\documentclass[pra,twocolumn,superscriptaddress,showpacs,nofootinbib]{revtex4}

\usepackage{graphicx}

\usepackage{bm}
\usepackage{fleqn}
\usepackage{amssymb}
\usepackage{epstopdf}
\usepackage{auto-pst-pdf}
\usepackage{amsmath}
\usepackage{amstext}
\usepackage{latexsym}
\usepackage{hyperref}
\usepackage{pstricks}
\usepackage{ulem}

\usepackage{inputenc}

\newcommand{\ket}[1]{\left\vert#1\right\rangle}

\newcommand{\beq}{\begin{equation}}
\newcommand{\eeq}{\end{equation}}

\begin{document}

\title{Scaling of the entanglement spectrum near quantum phase transitions}
\author{L. Lepori}
\affiliation{Departament de F\'{i}sica,
Universitat Aut\`{o}noma de Barcelona, E-08193 Bellaterra, Spain.}
\author{G. De Chiara}
\affiliation{Centre for Theoretical Atomic, Molecular and Optical Physics, School of Mathematics and Physics, QueenÕs University Belfast, Belfast BT7 1NN, United Kingdom.}
\author{A. Sanpera}
\affiliation{ICREA, Instituci\`o Catalana de Recerca i Estudis Avan\c{c}ats, E-08010 Barcelona, Spain.}
\affiliation{Departament de F\'{i}sica,
Universitat Aut\`{o}noma de Barcelona, E-08193 Bellaterra, Spain.}

\begin{abstract}
The entanglement spectrum describing quantum correlations in many-body systems has been recently recognized as a key tool to characterize different quantum phases, including topological ones. Here we derive its analytically scaling properties in the vicinity of some integrable quantum phase transitions and extend our studies also to non integrable quantum phase transitions in one dimensional spin models numerically. Our analysis shows that, in all studied cases, the scaling of the difference between the two largest non degenerate Schmidt eigenvalues yields with good accuracy critical points and mass scaling exponents.
\end{abstract}

\maketitle

The frontiers of condensed matter physics have broadened substantially beyond the paradigms of the twentieth century,  namely conventional superconductivity (BCS) and the band theory, with the discovery of topological quantum phases.  These novel states are widely thought to have great potential for technological applications. Indeed, first attempts in this direction are already in progress. Some examples are the use of the edge states for topological quantum computations \cite{Nay08} or the cheap synthesis of high-Tc superconductors via the doping of certain spin liquids \cite{Nag06}. However, for the sake of applications a deeper theoretical comprehension is required. Yet another milestone has been the gradual convergence of condensed matter physics and quantum information theory (QIT) motivated by the fact that quantum many-body systems 
are a natural territory of quantum entanglement, which is the basic computational resource in QIT. 
Such theoretical developments run in parallel and are stimulated by the spectacular experimental progress achieved in the areas of ultracold physics and condensed matter. New platforms such as ion traps, optical lattices, bosonic and fermionic atomic degenerate gases \cite{Annabook}, new superconducting materials, exotic quantum Hall systems are intensively investigated as quantum simulators. On these devices a huge set of new phases is postulated to be synthesizable.
In the light of these new exciting developments, the problem of how to characterize, detect and simulate new quantum phases of matter remains as one of the greatest challenges in the field.\\
Traditionally, the characterization of quantum phases and their phase transitions has been based on 
local order parameters $Q$, and the response to linear perturbations given by low order correlators e.g. $\langle S_{i} S_{j}\rangle$, where $S_{i}$ is an observable of site $i$ in a lattice system. 
Such description is in accordance to the ``standard" Ginzburg-Landau scenario of phase transitions. 
Here the order, which is associated to the breaking of some symmetry,  
is manifested by a change in the expectation value of a {\it{local}} order parameter $Q$.
However, there are important cases where an order parameter is not available, as in presence of topological phases or more simply whenever a local parameter is intrinsically difficult to construct or measure. Notable examples arise for instance in the presence of deconfined criticality, like at the transition semimetal-insulator on the graphene or in the Haldane-Shastry model \cite{Senth04, Bern04}, or in long range interacting systems \cite{dipolar}.  A similarly challenging situation realizes whenever the phase diagram of the material under investigation is not known at all, as for some high-Tc superconductors or for various lattice tight-binding models  (see for instance \cite{Meng10}). In all these cases a major problem is to identify a quantity able to detect a certain phase or phase transition.\\
In recent years, concepts from QIT have helped to understand the structure of quantum many-body systems and the computational power needed to simulate them \cite{Schuch07}. 
A major step in this direction was the study on how the bipartite entanglement of a block $A$ of a many-body system, as measured by the entanglement entropy $S=-\rho_{A}\log\rho_{A}$ ($\rho_A$ being the reduced density matrix), scales with the size $n_{A}$ of the block \cite{Vidal03}.
Later on, it was rigorously proven \cite{Hastings07} that for all 1D gapped quantum systems described by short range Hamiltonians, the entanglement entropy saturates to a constant independently of the size of the block. This behavior of the entropy, encountered in several areas of physics, is termed ``area law" \cite{Eisert10}. 
At the quantum phase transition (QPT) instead the entanglement entropy diverges logarithmically with the block size as $S\sim c\log(n_A)$ \cite{Holzhey94,Korepin04,Calabrese04} where $c$ is the central charge of the conformal field theory (CFT) describing the critical point. 
Out of criticality, the many-body system is not any longer entirely constricted by the central charge but has a dependence on the specific perturbations tuned to move from the transition point. It is in this regime that one expects the entanglement spectrum, i.e. the set of the eigenvalues $\{\lambda_{i}\}$ of the reduced density matrix $\rho_{A}$, to contain information that is not included in the entanglement entropy, being this quantity a single number. Indeed Li and Haldane \cite{Li08} and immediately later various authors \cite{Pollmann10} pointed out that the entanglement spectrum of topological systems, after a cut in a suitable Hilbert space, shows peculiar features in degeneracy, gap, distribution and it can be a valuable tool for the investigation of many-body systems. 
The structure of the entanglement spectrum reflects some specific features of the phase in which the system is, like symmetries, edge states, etc.
In view of these developments, we have recently examined \cite{DeChiara_Lepori} the entanglement spectrum 
focusing on its specific scaling features near some QPTs. In particular, using finite size scaling (FSS) we have numerically demonstrated that in the longitudinal and transversal Ising model, the Schmidt gap, i.e., the difference between the two largest non trivially degenerated eigenvalues of the reduced density matrix $\Delta\lambda=\lambda_{1}-\lambda_{2}$, correctly signals the critical point and scales with critical exponents related to the  CFT describing the transition point itself and to its specific perturbations. Recent studies in 2D systems also show the scaling of the entanglement spectrum near phase transitions \cite{Lauchli13,James12}.\\
Here, we extend our previous results first by deriving analytically the scaling of the Schmidt gap for spin-1/2 Ising model with transverse or longitudinal magnetic field.
Second, we analyze numerically the scaling properties of the entanglement spectrum in non integrable cases. 
Our main result is that for both, integrable and non-integrable, models the Schmidt gap always shows scaling behavior and the FSS of this quantity yields the mass scaling exponent and the precise critical point. \\
We stress that for a lattice system there are two different concepts of integrability: the first one concerns {\it lattice} integrability, equivalent to the total solvability of the lattice model (for details see \cite{Baxter,Mussardo}), while the second one involves the solvability of the infrared continuos theory 
describing the low energy sector of it
(the prominent one close to criticality) \cite{Mussardo}.  When not explicitly specified, in the following we will mean the second type of integrability.\\
The paper is organized as follows. In Section \ref{IsingCFT}, we first review previous  results concerning the entanglement spectrum and the Schmidt gap at criticality for finite size systems. Then, we derive analytically the closing of the Schmidt gap in the scaling regime (out of criticality) using the Baxter and Cardy's conjecture that for integrable models near criticality one can connect the entanglement spectrum to the characters of the conformal field theory (CFT) describing the critical point \cite{Baxter,Peschel_first,Peschel, Peschel2,conj,Cardy}. Finally, we compare the  derived analytical expressions with the results obtained by solving the spin-1/2 Ising chain in the transverse and longitudinal models. Notice that these two phase transitions share the same critical point but correspond to different universality classes. This is clearly reflected in values of  the scaling exponents obtained for them. In Section \ref{sec:spin1} we focus on more complex models, including also non integrable ones. Indeed integrability is known to constrain greatly the dynamics of a system \cite{Mussardo} and it could be conjectured to be responsible of the peculiar behavior of the Schmidt gap previously described. However, we observe that this is not the case. To this aim, using the density matrix renormalization group (DMRG) algorithm and the FSS \cite{dmrg,Fisher72}, we focus in particular on the spin-1 Heisenberg model with an uniaxial anisotropy  $D$ \cite{DeChiara11, Rodriguez10} which exhibits a Haldane topological phase surrounded by other quantum phases (N\'eel, dimer, large D and critical) belonging to very different universality classes. 
The Haldane-N\'eel transition is in  the transverse Ising universality class, then the closing of the Schmidt gap reproduces the results for the spin-$1/2$ transverse Ising model. 
 We discuss then the scaling properties in Haldane-dimer QPT described by the Wess-Zumino-Witten $SU(2)_{2}$ conformal theory. Finally, we analyze the Haldane-large~D transition where the critical exponents change along the critical line (having central charge $c=1$) and the N\'eel-dimer phase transitions, where the central charge changes continuously along the transition line. Using FSS techniques, for all the above transitions we derive the mass scaling exponent and the critical point. Finally, in Section \ref{sec:conclusion} we summarize our results and list some open problems. 

\section{The Schmidt gap in integrable models: The spin-$\frac{1}{2}$ Ising chains}
\label{IsingCFT}
The entanglement spectrum is the set of the eigenvalues of the reduced density matrix, known as  Schmidt eigenvalues, for any of the two blocks obtained after a bipartite splitting of a system in real space. For translational invariant systems and sufficiently large blocks, the point at which the bipartition is done is irrelevant, for simplicity we assume to be in the middle of the chain \cite{note2}.
After the bipartite splitting, one can always Schmidt decompose any pure state, e.g. the ground state, of the entire many-body system as
\beq
\ket{\psi_{GS}}=\sum_i\sqrt{\lambda_i} \ket{\phi_i^L}\otimes
\ket{\phi_i^R}, \label{schmidt}
\eeq 
where $\lambda_i\geq 0$ are the Schmidt coefficients with respect to the partition sorted in decreasing order, while $\ket{\phi_i^R}$ and $\ket{\phi_i^L}$
are the Schmidt eigenvectors  corresponding to the right (left) subsystem.\\
The starting point in deriving the entanglement spectrum is the fact that the reduced density matrix $\rho_R$ of a semichain can be expressed as the n-th power of the corner transfer matrix $A$ \cite{Baxter} constructed on a certain portion of the classical 2D lattice associated to the quantum chain: $\rho_R = A^n$ \cite{Cardy,Peschel_first}. For a square lattice, one can manage the construction such to have $n = 4$. If the lattice is integrable, $A$ can be in turn expressed as $A = \mathrm{exp}[- \tilde{H} ]$, where $\tilde{H}$ is a fictitious non local Hamiltonian defined on a semi-chain \cite{Cardy,Peschel_first}.  This property also holds for systems that are integrable only in their continuos infrared formulation \cite{conj}.  The eigenvalues of $\tilde{H}$, $\xi_k$, are related to the Schmidt eigenvalues $\lambda_k$ as follows: $\lambda_k = C e^{- n \, \xi_k}$, where $C$ is an overall constant such to fulfill the constraint $\sum_k \lambda_k = 1$. Clearly, the Hamiltonian $\tilde{H}$ must be gapless 
whenever $\rho_R$ is calculated for a given system at a critical point. Consequently, in this point $\{ \xi_k \}$ and $\{ \lambda_k \}$ must tend to a continuous distribution having statistics as in (\ref{rho_crit.}) below, therefore  the Schmidt gap must tend to zero.  We believe this property to hold also for nonintegrable models, since they are continuously related to integrable ones in the space of the Hamiltonian parameters \cite{Mussardo} whenever they share the same critical point.\\
Let us start by analyzing first the Schmidt gap at criticality for finite size systems. In the CFT framework, 
the reduced density matrix of semichain having size ${\ell}$ 
 can be expressed as \cite{Holzhey94,orus}: 
\beq 
\rho_{R}=\frac{1}{Z_{R}(q)}q^{-c/24} q^{L_{0}} \, ,
\label{rho_crit.}
\eeq
where $c$ is the central charge, $L_0$ is the zero-generator of the chiral Virasoro algebra, $Z_R(q)=\textrm{Tr} \,q^{L_{0}}
$ 
is the partition function of the subsystem with size $\ell$ of a torus (equivalent to a cylinder) \cite{Holzhey94, orus} and 
$q=\mathrm{exp}(i 2\pi \tau)$ is the modular parameter. The relation $\tau=i\kappa/\log(\ell/\eta)$ holds, where $\eta$ is a regularization cutoff and $\kappa$ is a positive constant. The Schmidt gap $\Delta\lambda$ can be expressed as:
\beq
\Delta\lambda(\ell) = \lambda_{1}-\lambda_{2}=\frac{1-q^{\alpha_1}}{Z_{R}(q)}= \frac{1-q^{\alpha_{1}}}{\ell^{c/12}} \, ,
\label{schmidtgapCFT}
\eeq
where we exploited the fact that the largest eigenvalue of the reduced density matrix is directly related to the single copy entanglement $S_{1}=-\ln\lambda_{1}$ \cite{singlecopy} and goes as $\lambda_1\sim \ell^{-c/12}$.
The coefficients $\alpha_{k} > 0$  have the form $\Delta + N$ $(N>0)$ where $\{\Delta\} $ are the scaling dimensions of the primary operators of the conformal theory describing the critical point and $\{N\}$
denotes the set of integer positive numbers labeling the conformal levels. Each level has a computable degeneracy $d_{N}^{(\Delta)}$. The numbers  $\Delta + N$, defining an ordering labeled by the index $k$ above, correspond to the eigenvalues of the operator $L_{0}$ \cite{Dif,Cardy}. Since the eigenvalues of this operator are fixed by the central charge so is the Schmidt gap at criticality for a finite size chain. \\
 Out of criticality, but still in its vicinity, the entanglement spectrum is expected to acquire specific features, loosing the universal dependence on the central charge valid at the critical point.
It can be inferred that for an integrable model, 
the trace of the $n$th-power of reduced density matrix $\rho_R$ for the semichain described {{\it close to criticality}} \cite{Mussardo} can be expressed by a character sum as \cite{Baxter, Cardy, conj}:
 \beq
\mathrm{Tr} \, \rho_R^n =  \frac{\sum_{\Delta} a_{\Delta}(q) \, \chi_{\Delta}  (q^n)}{\left[\sum_{\Delta} a_{\Delta}(q) \, \chi_{\Delta}(q)\right]^n} \,,
\label{char}
 \eeq 
where $\chi_{\Delta}(q)$ are the conformal characters  \cite{Cardy,Dif}:
\begin{equation}
\chi_{\Delta}(q) \equiv q^{-c/24+\Delta}\sum_{N = 0}^\infty d_{N}^{(\Delta)} \, q^{N}\,.
\label{eq:chi}
\end{equation}
Indeed, in these models the eigenvalues of $A^{4}$ are all of the form $q^{a N + b}$ and with computable degeneracies $d_{N}^{(\Delta)}$.
The parameter $q$, depending now on the correlation length $\xi$, tends to 1 at criticality ($\xi \to \infty$), there it agrees with the modular parameter $q=\mathrm{exp}[- 2\pi \kappa/\log(\ell/\eta)]$, appearing in (\ref{rho_crit.}), in the limit $\ell \to \infty$.

This observation allows us to express the Schmidt coefficients near criticality as: 
\beq
\lambda_k = \frac{a_{\Delta}(q) \, q^{- \frac{c}{24} + \Delta + N}}{\sum_{\Delta} a_{\Delta}(q) \, \chi_{\Delta}  (q)} = \frac{a_{\Delta}(q) \, q^{- \frac{c}{24} + \Delta + N}}{Z_R (q)} \, ,
\label{charsing}
\eeq 
$Z_R (q)$ is  added as a trace normalization for the $\{\lambda_k\}$.
The difficulty in evaluating the above expression comes from the lack of knowledge of the coefficients $a_{\Delta}(q)$ appearing in (\ref{char}) whose values are only known at criticality where they fulfill $a_{\Delta}(q=1)=1$. This limit can be inferred from the fact that $\mathrm{Tr} \, \rho_R^n$ is expressible at criticality just as a simple  sum of the characters, with all  coefficients equal to 1 \cite{orus,DeChiara_Lepori}.  For continuity the $a_{\Delta}(q)$, when nonzero, should be close to $1$ in the scaling limit. For  sake of generality we assume that they all have  a $q$-dependence.  Actually, the fact that on a certain side of a given phase transition some $a(q)$ vanish for every $q$ seems to suggest that each $a(q)$ generally assumes only values 0 or 1, probably due to some protecting symmetries, the same ones underlying integrability \cite{Sierra}. We leave this issue as an open point. Finally, notice that the positivity of $\lambda_k$ implies positiveness also of $a_{\Delta}(q)$.
 
\subsection{Transverse magnetic field Ising Model}
\label{transvCFT}
We will go through the detailed calculation for a known case, the spin-1/2 transverse Ising model, having  central charge $c = \frac{1}{2}$ at the critical point. Obviously, the same discussion holds for all the lattice models that at criticality belong to the same universality class. 
A direct analysis of the entanglement spectrum obtained numerically (see subsection \ref{sectnum})  shows that 
in the low-temperature  (ferromagnetic) phase the character entering in (\ref{char}) is $\chi_{\frac{1}{16}} (q)$ (indeed the entanglement
spectrum is doubly degenerate) while in the high-temperature (paramagnetic) phase $\chi_{0} (q)$ and $\chi_{\frac{1}{2}} (q)$ are relevant. 
We consider the second case, since it also reproduces the Ising transition in the spin-1 model approaching the Haldane phase from the N\'eel one.
Setting $n=1$  in (\ref{char}), the Schmidt gap $\Delta\lambda \equiv \lambda_{1} - \lambda_{2}$ can be written as:
\beq
\Delta\lambda =\frac{q^{-\frac{c}{24}} \left[a_0 (q) - a_{\frac{1}{2}} (q) \, q^{\frac{1}{2}}\right]}{Z_R(q)} = \frac{\left [1- \beta_{0 ,\frac{1}{2}} (q) \, q^{\frac{1}{2}} \right] }{\xi^{\frac{c}{12}}} \, ,
\label{outCFT} 
\eeq 
where
$\beta_{0, \frac{1}{2}}(q) =  a_{1/2}(q)/a_{0} (q) $ and we have that 
 $S_{1} = - \text{ln}  \lambda_1 \approx \frac{c}{12} \, \text{ln} \, \xi$. 
In this particular model $a_0(q)$ and $a_{\frac{1}{2}}(q)$ can be probably computed expressing $\rho_R$ as a thermal ensemble of free fermions \cite{Peschel2009}, however we will see that the precise expressions for them are not strictly required in our calculation.\\
We want to evaluate $q$ in terms of $\xi$ in the limit $q\to1$, i.e. close to the critical
point, where unfortunately the series in Eq.~(\ref{eq:chi}) it is not so useful since it must be re-summed before taking the limit $q \to 1$. However, the
Virasoro characters transform linearly under a modular
transformation $q\to\tilde q$:
\begin{equation}
\chi_{\Delta}(q)=\sum_{\Delta'}S_\Delta^{\Delta'}\chi_{\Delta'}(\tilde
q)\,,
\label{modtransf}
\end{equation}
where $q$ and
$\tilde q$ are linked each others by the standard relation \cite{Cardy, conj}:
\beq 
\mathrm{ln} q \, \mathrm{ln} \tilde{q} = 4 \pi^2 \, .
\label{relqq}
\eeq
The series  (\ref{eq:chi}) expressed in $\tilde{q}$ has the virtue that the critical limit $\tilde{q} \to 0 $ can be taken before to resum all the terms \cite{conj}.
$S_\Delta^{\Delta'}$ is called modular matrix  and 
characterizes the modular transformations on the torus  for a given CFT \cite{Cardy}. 
For the Ising CFT  $(c = \frac{1}{2})$ it reads \cite{Cardy}:
\beq
\left(
\begin{array}{ccc}
1 & 1 & \sqrt{2}\\
1 & 1 & -\sqrt{2}\\
\sqrt{2} & -\sqrt{2} & 0
\end{array}
\right)
\label{modising}
\eeq
in the basis $(\chi_0 (q), \, \chi_{\frac{1}{2}} (q), \, \chi_{\frac{1}{16}} (q))^{T}$ or equivalently $(\tilde{\chi}_0 (q), \, \tilde{\chi}_{\frac{1}{2}} (q), \, \tilde{\chi}_{\frac{1}{16}} (q))^{T}$ (the matrix is nilpotent of order two).\\
Since the first column, concerning  the expression of the $\chi_{i} (q)$ in terms of $\chi_0 (\tilde{q})$, contains only positive numbers, as well as the coefficients $a_{\Delta}(q)$ in (\ref{char}), it is clear that, after the transformation (\ref{modtransf}), the expression in $\tilde{q}$ for $\mathrm{Tr} \, \rho_R^n$, analogous to (\ref{char}), surely contains the character $\chi_0 (\tilde{q})$. Notice that this claim does not depend strictly on the particular expression (\ref{char}).
We consider now the functional dependence of $\mathrm{Tr} \,\rho_{R}^n $  in terms of $\tilde{q}$. It is straightforward to extract
the leading behavior in the limit $\tilde{q}\to 0$, as well as the first corrections. The leading term
comes from the character $\chi_0 (\tilde{q})$ (surely present in  $\mathrm{Tr} \, \rho_R^n (\tilde{q})$):
\beq
\mathrm{Tr} \,\rho_{R}^n\sim \left[a_0(q) S_0^0 + a_{\frac{1}{2}}(q) S_{\frac{1}{2}}^0\right]^{1-n}\left({\tilde
q}^{-c/24}\right)^{(\frac{1}{n} - n)} \,,
\label{exp}
\eeq
which can be used to obtain the Renyi entropy  
\begin{eqnarray}\label{eqSnmod}
S_n&\equiv& \frac{1}{1-n}\mathrm{Tr} \, \rho_{R}^n 
 \\
&\sim& -\frac c{24}\left(1+\frac1n\right)\ln\tilde
q+\ln \left[a_0(q) S_0^0 + a_{\frac{1}{2}}(q) S_{\frac{1}{2}}^0\right].
\nonumber
\end{eqnarray}
Comparing  (\ref{eqSnmod}) with the leading term for the Renyi entropy out of criticality \cite{Calabrese04, conj}
\beq
S_n \simeq \frac{c}{12}\left(1+\frac1n \right)\ln \xi ,
\label{renyi}
\eeq
one arrives to  $\tilde
q = \eta \, \xi^{-2}$ ($\eta$ being a constant) and $q = \exp\left[\frac{ 4 \pi^2}{\mathrm{ln} \eta - 2 \mathrm{ln} \xi}\right]$. \\
Finally, the term $g^2 = \ln \left[a_0(q) S_0^0 + a_{\frac{1}{2}}(q) S_{\frac{1}{2}}^0\right]$ is related to the
Affleck-Ludwig boundary entropy $g$ \cite{AfLud}.
Notably a measure of it can give direct information on $a_0(q)$ and $a_{\frac{1}{2}}(q)$. Since $a_{\Delta}(1) = 1$, $g^2$ is expected to be well approximated by $g^ 2 \approx \ln \Big(S_0^0 + S_{\frac{1}{2}}^0\Big)$.\\	
Inserting the expression for $q$ in (\ref{outCFT})
and expanding the exponential yields:
\beq
\begin{array}{c}
\Delta \lambda (\xi) 
=  \frac{c (q)}{\xi^{\frac{c}{12}}}  - \frac{2 \pi^2}{\xi^{\frac{c}{12}} (\mathrm{ln} \, \eta -2 \, \mathrm{ln} \xi)} =
\frac{\tilde{c} (\xi)}{\xi^{\frac{c}{12}}}  - \frac{2 \pi^2}{\xi^{\frac{c}{12}} (\mathrm{ln} \, \eta -2 \, \mathrm{ln} \xi)}
\label{final} 
\end{array}
\eeq 
where $c(q) = 1 - \beta_{0 ,\frac{1}{2}} (q)  $.
The limit $\tilde{c}(\xi) \to 0$ is not trivial and one cannot  naively approximate  $\tilde{c}(\xi) \approx 0$, the peculiar functional form of $\tilde{c}(\xi)$ determines instead which term in the r.h.s of (\ref{final}) is dominant for $\xi \to \infty$ (criticality).\\
Symmetry arguments discussed at the end of the last subsection and general arguments from the standard scaling theory (we take into account that no marginal operators are present)~\cite{Cardybook}
lead us to believe that $\tilde{c}(\xi) \to 0$ more rapidly than $\xi^{-1}$. Moreover, if our conjecture that $a_{\Delta}(q)$ are always either 0 or 1 is true, this term is exactly vanishing.
Using the scaling relation $\xi = \frac{1}{4 \pi} \, |g- g_c|^{- 1} $ and $c = \frac{1}{2}$,
 we arrive to the final expression:
\beq \Delta \lambda (|g-g_c|) =   A \, \frac{|g-g_c|^{\frac{1}{24}}}{B - \mathrm{ln} |g-g_c|}   \, ,
\label{finalg}
\eeq
with $A = \pi^2 (4 \pi)^{\frac{1}{24}} \approx 10.97$. The value for $B$ instead is not simply derivable.
The factor $\frac{1}{4 \pi}$, required for the estimation of $A$, can be derived from the exponential decay rate of the two-point correlation function of the energy density operator \cite{Mussardo} and considering that the mass gap scales as: $M = 2 \pi \, |g-g_c|$ \cite{masscoupling}.
Notice that in (\ref{finalg}), the dependence on the precise characters content  
is only included in the parameters $A$ and $B$.  The relation (\ref{final}), in particular, is expected to be universal (but with varying $c(q)$ and $b_{0, \frac{1}{2}}(q)$) for integrable models.\\
Formula (\ref{finalg}) suggests that the scaling of the Schmidt gap depends mainly  (in its functional form) on the central charge and on the correlation length of the system (in turn depending on the Hamiltonian operator perturbing the critical point), ingredients already encoded in (\ref{renyi}). The precise expressions for $A$ and $B$ depend instead on the Schmidt eigenvalues, via the parameters $a_{\Delta}(q)$. However the derivation presented in this section works only for integrable perturbations and we leave as an open question whether in a general non-integrable case a stronger dependence on the full entanglement spectrum holds.\\
Notice that for $g \to g_c$ the expression (\ref{finalg}) for the Schmidt gap tends to:
\beq
\Delta \lambda (|g-g_c|) = - A \, \frac{|g-g_c|^{\frac{1}{24}}}{\mathrm{ln} |g-g_c|}   \, ,
\label{finalgas}
\eeq 
to be considered the correct leading term in the thermodynamic limit. The factor $\frac{1}{\mathrm{ln}|g-g_c|}$ correcting the power law was already noticed in \cite{Calabrese08,conj}.\\
However we point out that, dealing with finite chains, the $B$ term in the denominator of Eq.~(\ref{finalg}) is required to be kept, since the quantity 
$\mathrm{ln} \, \xi$ in (\ref{final}) is prevented by the finite size to get sufficiently big in magnitude such to make $B$ negligible. For this reason the comparison with DMRG data in the next section will be performed assuming (\ref{finalg}).

\subsection{Longitudinal magnetic field Ising model}
\label{longCFT}
In this case 
the character content in (\ref{char}) 
can be guessed by simple considerations. In a generic case, the expression (\ref{char}) at the critical point contains all the characters of the CFT describing it, while outside the critical point only some of them appear on each side of the transition. 
It's then clear that the union of the sets of characters appearing in $\mathrm{Tr} \, \rho_R^n(q)$ on every side of the transition must be equal to the entire set of characters appearing in (\ref{char}) at criticality. The Ising model with longitudinal magnetic field displays a QPT between two phases with no residual symmetry, thus no constraints on the character content, directly related to the degeneracies of the Schmidt eigenvalues. The two phases have both a single vacuum, they share the same $S$-matrix and they can be mapped into each other \cite{Mussardo}.
For this reason we conjecture that, out of criticality,  $\mathrm{Tr} \, \rho_R^n(q)$  contains  all  three characters:
\beq
\mathrm{Tr} \, \rho_R^n(q)  = \frac{a_0 (q) \, \chi_{0}(q^n) + a_{\frac{1}{16}} (q) \, \chi_{\frac{1}{16}}(q^n) + a_{\frac{1}{2}} (q)  \, \chi_{\frac{1}{2}}(q^n)}{Z_R(q)^n}\, 
\label{ismag}
\eeq
where again $a_{\Delta} (q) > 0$. Notice however that in order  to obtain a Schmidt scaling law analogous to (\ref{finalg}), the explicit character content is not explicitly required, again thanks to the positivity of the coefficients $a_{\Delta}(q)$ and to the peculiarity of the matrix (\ref{modising}).
Setting again $n=1$ in (\ref{char}) and (\ref{ismag}) and taking into account (\ref{charsing}), the Schmidt gap can be expressed as:
\beq
\Delta\lambda =\frac{a_0 (q) - a_{\frac{1}{16}} (q) \, q^{\frac{1}{16}}}{Z_{R}(q)} 
\label{CFTlongi} 
\eeq
Using the same arguments and calculations as in the previous subsection and
the scaling relation $\xi = \alpha \, |g- g_c|^{- \nu} $, we obtain the formula:
\beq
\Delta \lambda (|g-g_c|) =   \tilde{A} \, \frac{|g-g_c|^{\frac{1}{45}}}{\tilde{B} - \mathrm{ln} |g-g_c|}   \, ,
\label{finalg2}
\eeq
with $\tilde{A} = 19.27$ (while again the value for $B$ is not easily derivable) and
in the thermodynamic limit
\beq
\Delta \lambda (|g-g_c|) =  - \tilde{A} \, \frac{|g-g_c|^{\frac{1}{45}}}{ \mathrm{ln} |g-g_c|}   \, .
\label{finalg2as}
\eeq
We explicitly inserted in (\ref{finalg2}) the values of the scaling exponents $\nu=8/15$ and $\alpha \approx 0.38$, this last values was estimated numerically in \cite{Ritt} from the decay rate of the two-point correlation function for the spin operator.\\
In (\ref{finalg2}) again the dependence on the precise characters content in (\ref{ismag}) is only in the parameters  $\tilde{A}$ and $\tilde{B}$.

\subsection{Comparison with numerical data} 
\label{sectnum}
\begin{figure}[h]
\begin{center}
\includegraphics[scale=0.50]{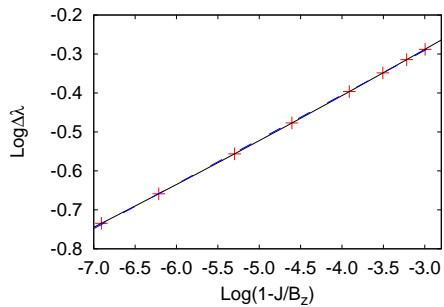}
\caption{Schmidt gap closing in the vicinity of the critical point for the transverse Ising model. Comparison between the analytical prediction from CFT (\ref{finalg}) (solid grey line), the power law scaling (\ref{powerlaw}) (blu dashed line)  and the numerical values (crosses) obtained by exact diagonalization. We consider a chain of 12000 sites. 
}
\label{Isingscalinga}
\end{center}
\end{figure}

\begin{figure}[h]
\begin{center}
\includegraphics[scale=0.50]{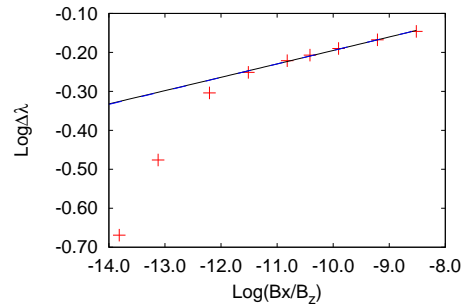}
\caption{Schmidt gap closing in the vicinity of the critical point for the longitudinal Ising model. Comparison between the analytical prediction from CFT (\ref{finalg2})  (solid grey lines), the power law scaling (\ref{powerlaw}) (blu dashed line), and the numerical values (crosses) obtained by solving the model with DMRG. We consider a chain of 
1536 sites. Notice that the fit is restricted to the six largest values.}
\label{Isingscalingb}
\end{center}
\end{figure}

At this point we want to check the predictions outside criticality obtained analytically against the results obtained by the scaling of the Schmidt gap when solving exactly or numerically the corresponding spin model:
\begin{equation}
\label{eq:Isingmodel}
H_{Ising}=-J\sum_i \sigma_x^i\sigma_x^{i+1} -B_x\sum_i \sigma_x^i-B_z\sum_i \sigma_z^i \, .
\end{equation}
The model (\ref{eq:Isingmodel}) reduces (i) for $B_{x}=0$ to the transverse Ising model, that is critical  at $J=B_z$, and  (ii) to the longitudinal Ising model for $B_{x}/B_{z} \neq 0$; both models are described at criticality by a CFT with central charge $c=1/2$, but they are in  different universality classes \cite{Cardy,Mussardo}. The entanglement spectrum and the Schmidt gap for both models were investigated in \cite{DeChiara_Lepori}. For the transverse Ising model the entanglement spectrum can be computed employing the Jordan-Wigner and Bogoliubov transformations to map the model into  a system of non interacting fermions, so that  the Schmidt eigenvalues can be straightforwardly extracted \cite{Peschel2009}.
The Ising model with a longitudinal field perturbation is not integrable in the lattice and therefore has to be solved numerically, by means, e.g. of DMRG at $J=B_z$ and for different values of $B_x/B_z \neq 0$. This model is however integrable in its infrared continuous limit, as well as the transverse Ising chain.
In both cases we found~\cite{DeChiara_Lepori} that, when approaching the critical point,  
the Schmidt gap displays scaling behavior with the size of the system $\ell$ and it allows for a FSS analysis, as for a standard local order parameters in a second order phase transition ~\cite{Fisher72}:
\begin{equation}
\Delta \lambda(\ell, g) \simeq \ell^{-\beta_{\Delta\lambda}/\nu} f_{\Delta\lambda}\left(|g-g_c| \ell^{1/\nu} \right) \, .
\label{powerfit}
\end{equation}
The quantity $\nu$ characterizes the divergence of the correlation length, while $\beta_{\Delta\lambda}$ plays the role of an ``order parameter" exponent. Indeed, the ansatz (\ref{powerfit}) implies a power-law hypothesis for the Schmidt gap  in the infinite limit:
\beq 
\Delta\lambda = k \, |g-g_c|^{\beta_{\Delta\lambda}} \, .
\label{powerlaw}
\eeq
While for local order parameters as the magnetization this hypothesis is under control  (see note \cite{note3}), for the Schmidt gap, being a non local quantity, the justification is highly non trivial, 
and it must be checked a posteriori, for instance probing the correctness of the $\nu$ exponent obtained from (\ref{powerfit}). \\
From the scaling of Schmidt gap by (\ref{powerfit}) we extract the following results (see \cite{DeChiara_Lepori}): for the transverse Ising model the critical point is located at $J/B_z =1$ and the critical exponents $\beta_{\Delta\lambda}=0.124 \pm 0.002$ and $\nu=1.00\pm 0.01$. These values agree very closely with the critical exponents on the Ising transition, suggesting that the Schmidt gap in the vicinity of the critical point correctly signals the location of it and it scales universally with critical exponents related to the CFT describing the transition. 
For the longitudinal Ising model the scaling of the Schmidt gap
leads to the critical exponents: $\nu= 0.50\pm0.05$ and $\beta_{\Delta\lambda}=0.055\pm0.005$, which again are in good agreement with the exact values of the corresponding exponents for the mass gap $(\nu=8/15)$ and for the magnetization order parameter ($1/15$).\\
To test the versatility of our estimates we compare them with the numerical results for the Schmidt gap in the transverse  and longitudinal Ising model with fixed system size.
Surprisingly enough, the scaling behavior obtained from the power law ansatz (\ref{powerlaw}) and from the expressions (\ref{finalg}) and (\ref{finalg2})
 are strongly compatible, as displayed in  Figs. (\ref{Isingscalinga}) and (\ref{Isingscalingb}). 	\\
 For the transverse Ising, the fitted parameters in (\ref{powerlaw}) are 
$\beta_{\Delta\lambda} = 0.114$ and $k = 1.05$.  
Assuming instead (\ref{finalg}), the fit yields
$B = 8.94$ and  $A =  10.27$, to be compared with the expected value $A = 10.97$. 
The succesful comparison between numerical data and the derived scaling law  (\ref{finalg})
represents a further argument in favor of the Baxter and Cardy's conjecture \cite{Baxter, Cardy, conj}.
Moreover it supports the claim that for finite size systems the $B$ term in (\ref{finalg}) is needed.\\
For the longitudinal Ising model, the power-law fit (\ref{powerlaw}) of the numerical data for a chain of 1536 sites gives
 $\beta_{\Delta\lambda} = 0.03$ and $k = 1.16$;
assuming instead  (\ref{finalg2}) we obtain $\tilde{B} = 71$ and  $A = 84.2$, to be compared with a theoretically expected value $A = 19.27$.   
Compared with the transverse Ising model, the larger error between the measured and the expected values of $A$ is probably due to the smaller size of the analyzed chain and to the bigger sensitivity of it to finite size corrections.
\section{Schmidt gap in spin-1 chains: scaling in absence of integrability}
\label{sec:spin1}
For non-integrable models the conjecture used to derive the scaling of the Schmidt gap in the previous section does not hold, since the conformal degeneracies are spoiled here by the infrared relevant perturbations  (in the renormalization group (RG) sense) of the critical theory.  
One could expect that, even for non-integrable cases, very close to the critical point ($q \approx 1 $) or to an integrable transition this effect to be very small and the conjecture to be still  approximately valid. However this regime is probably too small to allow us a reliable study of the scaling of the Schmidt gap and the numerical analysis is the unique way to proceed.\\
In what follows we analyze the scaling properties of the entanglement spectrum in the various quantum phases appearing for the spin-1 chain in the presence of an uniaxial anisotropy $D$. Some of the cases here discussed are non-integrable to the best of our knowledge.
\begin{figure}[h!]
\begin{center}
\includegraphics[scale=0.40,clip=true]{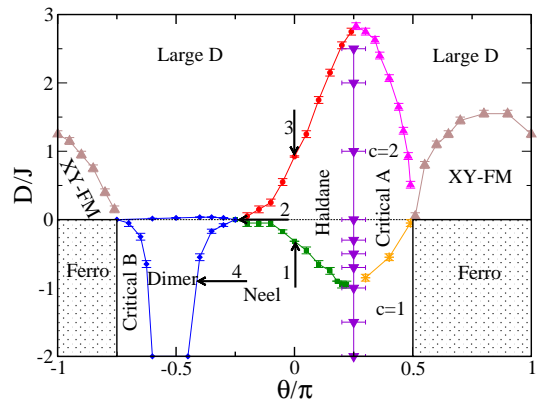}
\caption{Phase diagram of the spin-1 Hamiltonian (\ref{eq:H}) as a function of $\theta$ and $D$ computed
by DMRG. The details about the phases 
can be found in \cite{DeChiara11}.  We also put an arrow and a
corresponding number for the four critical points that we study more
extensively in the text.}
\label{fig:phase}
\end{center}
\end{figure}
The Hamiltonian describing a spin-1 chain is given by
\begin{equation}
\label{eq:H}
 H =\sum_i H_i(\theta) +D\sum_i S_{zi}^2,
\end{equation}
where $\bm{S}_i=(S_{xi},S_{yi},S_{zi})$ are the $i$-th site angular
momentum operators and $H_i(\theta)=\cos(\theta)
\bm{S}_i\cdot\bm{S}_{i+1}+ \sin(\theta)
(\bm{S}_i\cdot\bm{S}_{i+1})^2$. 
For $D=0$, the model is the well known 
bilinear-biquadratic spin-1 chain. The phase diagram for this chain as a
function of $\theta\in[-\pi;\pi]$  is known, see for example \cite{Schollwoeck96}. For $-\pi/4<\theta<\pi/4$, the system is in the
Haldane phase. At $\theta=-\pi/4 $ we obtain the so called  Takhtajan-Babujan spin-1 chain, a model integrable on the lattice. From the solution of the Bethe Ansatz equations, the only elementary excitations are known to be a doublet of gapless spin-1/2 spinons, with total spin 0 or 1 \cite{TB}. 
A detailed description of the phase diagram for $D \neq 0$ and its realization by ultra-cold gases is reported in \cite{Rodriguez10, DeChiara11}.
As shown in Fig.~\ref{fig:phase}, the uniaxial anisotropy leads to several phases 
surrounding the Haldane one. Those are: (i) the N\'eel phase, appearing for
negative $D$; (ii) the dimer phase, present for negative $D$, but
also for very small positive values of $D$; (iii) a large-$D$ phase whose ground state is connected adiabatically to the state in
which all spins have a zero angular momentum in the $z$-component; (iv) a critical gapless phase. Further to these phases, the model exhibits a ferromagnetic phase, an XY-ferromagnetic phase and a critical XY-antiferromagnetic.
To characterize the Haldane topological phase and to locate precisely its boundaries,
we compute the dimer order parameter $\mathcal D= H_i(\theta)-H_{i+1}(\theta)$, the staggered magnetization per site $M_z =1/L \sum_i (-1)^i S_{zi}$, acting as an order parameter for the N\'eel phase, and finally the entanglement spectrum.
Additionally, the Haldane phase is characterized by a non vanishing string order parameter \cite{Schollwoeck96}:
\begin{equation}
\label{eq:string}
O = \lim_{r\to\infty} \langle S_{zi}\exp[i\pi\sum_{j=i+1}^{i+r-1} S_{zj}]S_{zi+r}\rangle \, .
\end{equation}

\subsection{N\'eel-Haldane phase transition} 
\begin{figure}[h]
\begin{center}
\includegraphics[scale=0.27]{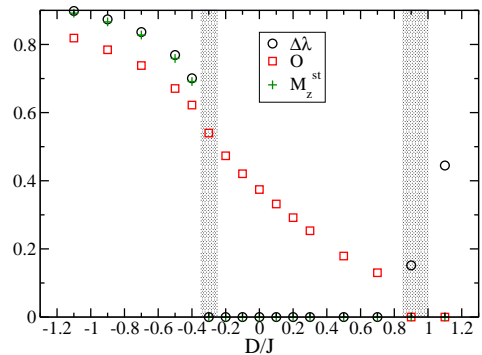}
\caption{ (Figure from \cite{DeChiara11}) The Schmidt gap, the string order parameter and the staggered magnetization for the Hamiltonian (\ref{eq:H}) on the $\theta = 0$ line,  shown as a function of $D/J$. 
The shaded area represents the scaling regions close to the critical points where our numerical analysis is less reliable.}  
\label{fig:hal-neel}
\end{center}
\end{figure}

As a first step we concentrate our analysis on the  N\'eel-Haldane phase transition along $\theta=0$, the uniaxial anisotropy $D$ is now the control parameter, as schematically shown in Fig. \ref{fig:phase} with arrow 1. In our DMRG simulations we apply a small magnetic field to the first spin to select one of the two degenerate ground states.
The FSS analysis for the staggered magnetization yields values for the location of the critical point and for the critical exponents respectively $ D_c = -0.315$, $\beta = 0.11$ and $\nu = 1.01$, very close to those obtained from the FSS of the Schmidt gap   $D_c = -0.315, \beta = 0.11,  \nu = 1.04 $. These results are comparable with the ones from the quantum MonteCarlo~\cite{Albuquerque09}  $D_c = -0.316, \beta = 0.147,  \nu = 1.01 $ and are in agreement with the conjecture  \cite{Albuquerque09} that this transition belongs to the transverse Ising universality class $\beta=0.125$, $\nu=1$. 
The $12\%$ discrepancy in our estimate of $\beta$  can be attributed to the relatively small size of the considered chains, the same for MonteCarlo results.
This similarity suggests us to compare the statistics of the Schmidt eigenvalues from DMRG with that one obtained by the character analysis performed out of criticality for the transverse Ising model in Sec. \ref{IsingCFT}. By direct inspection we checked that the statistics in the two cases perfectly agree, confirming that these two different models are in the same universality class and therefore share the same  entanglement spectrum outside, although still close to, criticality. \\
As for the transverse Ising model, the scaling of the Schmidt gap leads to similar
critical parameters as those found from the FSS of the staggered magnetization.
This correspondence seems to be closely related to the observation that, even far from the transition point, staggered magnetization and Schmidt gap behave almost identically, as shown in Fig. \ref{fig:hal-neel}.
\subsection{Dimer-Haldane phase transition}
\begin{figure}[h]
\begin{center}
\includegraphics[scale=0.65]{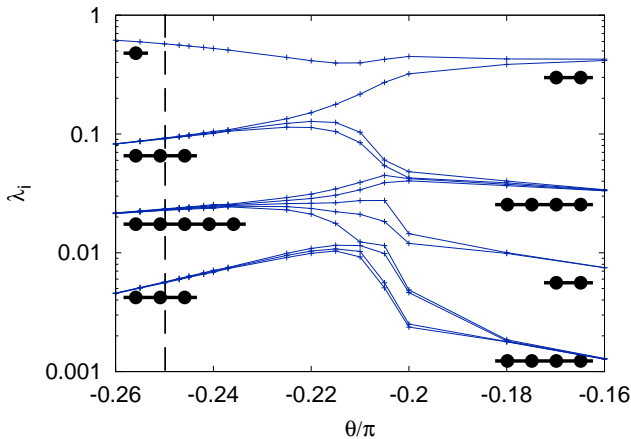}
\caption{Schmidt spectrum close to the Dimer-Haldane phase transition for $D=0$ as a function of $\theta$ and for $L=384$. The critical point is at $\theta/\pi = - 1/4$ and it is indicated by a vertical dashed line. The discrepancy between the location of the critical point and the zone where the statistics of the entanglement spectrum changes is a finite size effect. We indicate the degeneracies with small clusters of dots in the two phases. Lines joining the numerical data are only a guide to the eye.} 
\label{Hal-dim}
\end{center}
\end{figure}

 We analyze now the transition from Dimer to Haldane phase along $D=0$ and changing $\theta$ (see arrow $2$ in Fig.~\ref{fig:phase}). 
The conformal point related to this transition is the previously mentioned Takhtajan-Babujan point \cite{TB} and it is known to be described by a $SU(2)$ current algebra (Wess-Zumino-Witten, WZW) theory at level 2 (following the standard notation we will denote it as $SU(2)_2$). This theory is characterized by a central charge $c=3/2$ and two primary relevant operators with spin $1/2$ and $1$ respectively and conformal dimension $3/16$ and $1/2$ \cite{Cardy,Mussardo}. We focus on the FSS of the dimer order parameter $\mathcal D$ and the Schmidt gap $\Delta \lambda$. The lattice model (\ref{eq:H}) with $D = 0$ is $SU(2)$ invariant, therefore, provided that the cut in space does not break 
explicitly this symmetry \cite{note2}, we expect the Schmidt spectrum to organize in multiplets corresponding to the representations of this group. 
For every considered even size $L$, the cut in the middle point breaks an $SU(2)$-singlet either in the dimerized or in the Haldane phase, depending of the parity of $L/2$. However the effects of the $SU(2)$ breaking are negligible for long chains, as it usually happens for boundary effects, then at the end we expect the spectrum to be always arranged in multiplets, no matter the position of the cut. This picture is confirmed in Fig. \ref{Hal-dim}, where multiplets are clearly visible. 
We notice also the peculiar distribution of the Schmidt eigenvalues in both  the Haldane and the dimer  phases: the first one has only even multiplets,
while the second one contains only odd multiplets. This structure agrees with recent results obtained in \cite{Wen11,Cirac11}, where all the possible phases of the 1D gapped and short-range interacting Hamiltonians are classified by assuming a matrix product state (MPS) ansatz for them and studying the isometries of MPS. There a  correspondence was found with the different projective representations of the local invariance group $G$ for the analyzed energy eigenstate (generally the ground state). It is then not difficult to see that one can equivalently classify the linear representations of the covering group $\tilde{G}$ in subsets that are one-to-one related to  its center group (the precise relation comes directly from group representation theory, see \cite{Humph}).
The Hamiltonian (\ref{eq:H}) along the line $D=0$ is invariant under $SO(3)=SU(2)/Z_2$  therefore different phases corresponds to the two sets of linear representations of $SU(2)$ labeled by the different elements of its center $Z_2$: the spin integer and semi-integer representations. This link translates directly in the 
\begin{figure*}[htbp]
\begin{center}
\includegraphics[scale=0.35]{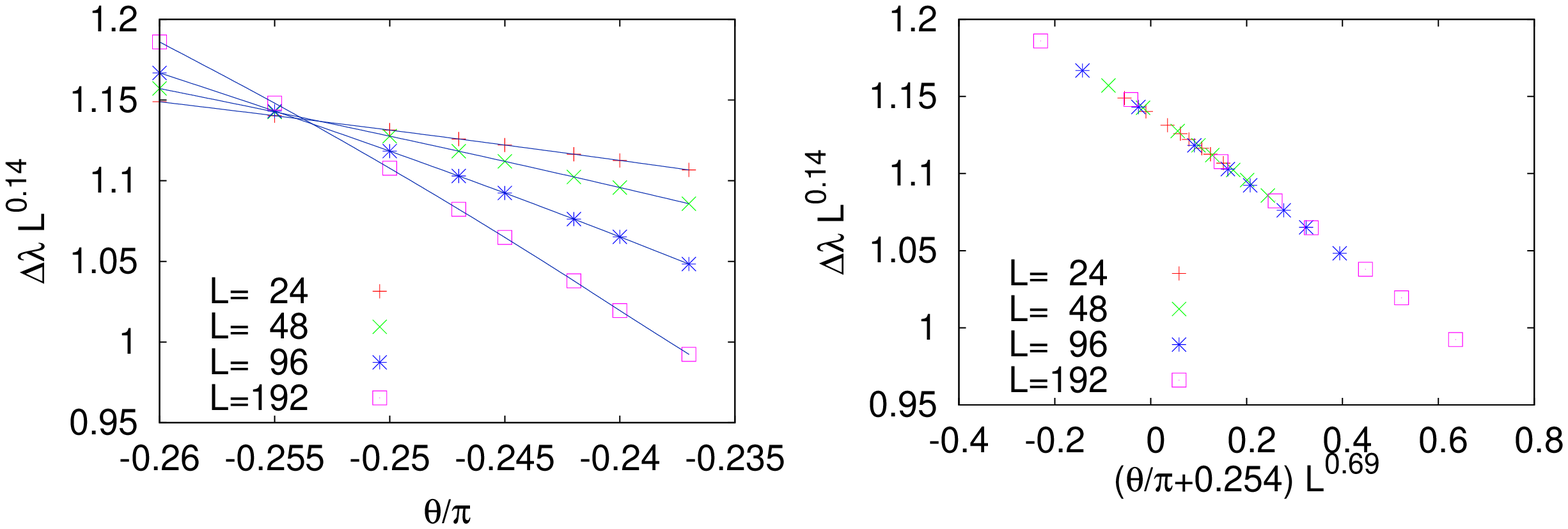}
\includegraphics[scale=0.35]{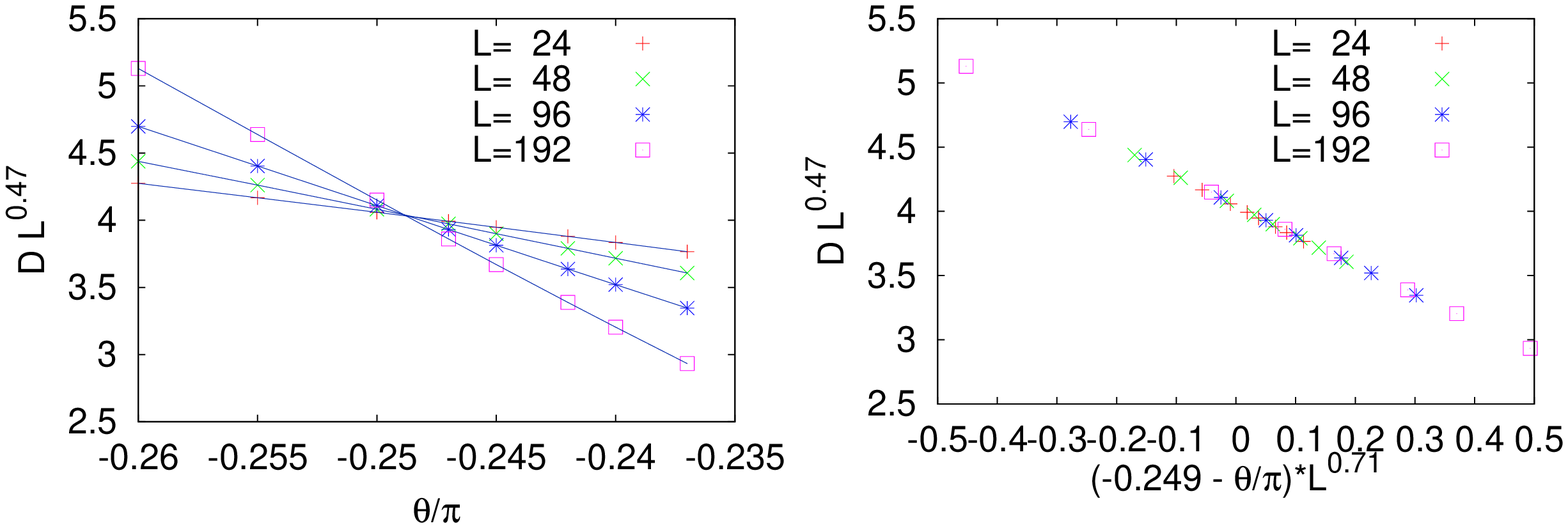}
\caption{Finite size scaling close to the Haldane-dimer transition. From the left size, the first and second panels are the FSS results for the Schmidt gap $\Delta\lambda$, while the third and fourth panels refer to the FSS of the dimer order parameter.}
\label{fig:dimerhaldanescaling}
\end{center}
\end{figure*}
content of multiplets of the Schmidt spectrum, as shown clearly in Fig.~\ref{Hal-dim}.\\
The FSS analysis for the dimer order parameter $\mathcal D$ and the Schmidt gap $\Delta\lambda$ is shown in Fig.~\ref{fig:dimerhaldanescaling}. The results for the critical point extracted from $\mathcal D$ and $\Delta\lambda$ are $\theta_c=-0.249\pi$ and $\theta_c=-0.254\pi$ respectively, very close to the exact result $\theta=-\pi/4$, confirming the effectiveness of the Schmidt gap as an order parameter signaling a phase transition. The estimates for the correlation length exponent are $\nu=1.41$ and $\nu=1.45$ respectively and they agree with the exponent calculated directly from the scaling of the energy gap: $\nu = 1.43$. Again the scaling law ansatz (\ref{powerfit}) turns out to be effective, yielding the correct value for $\nu$. Finally, the critical exponent for the dimer order parameter is $\beta=0.66$ 
and for the Schmidt gap is $\beta=0.20$.\\
Let us try to gain some insight on these exponents using arguments from CFT. The $D = 0 $ line has $SU(2)$ symmetry, therefore the most general perturbation of the critical point $SU(2)_2$ along this line is expected to be a combination of certain powers of the singlets constructed by the relevant fields of the $SU(2)_2$ WZW theory. The singlet constructed by the spin-$\frac{1}{2}$ multiplet is $(\phi_{\frac{1}{2}}^{R} \phi_{- \frac{1}{2}}^{L} - \phi_{- \frac{1}{2}}^{R} \phi_{\frac{1}{2}}^{L})$, while the singlet constructed by the spin-1 multiplet is  $(\phi_{0}^R \phi_{0}^L - \phi_{1}^{R} \phi_{-1}^{L} - \phi_{1}^{L} \phi_{-1}^{R})$. Notice that here $R$ and $L$ denote the chiralities of the conformal $SU(2)_2$ algebra \cite{Cardy, Mussardo}. The last possible singlet is the scalar product of the chiral and anti-chiral $SU(2)$ vectorial current $s_J = \vec{J}_R \cdot \vec{J}_L$. Eventually we obtain that the most general relevant perturbation along $D = 0$ is $\sum_n (s_{1/2})^n + (s_{1})^m + \sum_k (s_J)^k$. Each $(n,m,k)$-powers can be further decomposed in a sum of terms derived from the one point expansions (OPE) of the $(n, m, k)$-product of chiral fields \cite{Cardy}.  An explicit calculation employing bosonization \cite{2books,Affshort,Cardy} shows that in our case the Hamiltonian density of the continuous theory close to the critical point is
 \beq
 H =  H_{SU(2)_2} + a(\theta) \, s_1 + b \, s_J \, ,
 \label{HdimHal}
\eeq
where $a (\theta) \propto \theta - \frac{\pi}{4} $.  Interestingly, $s_1$ turns out to be a mass term for the three Majorana fermions describing the $SU(2)_2$ point. The application of both $s_1$ and $s_J$ was argued to give rise to a path from $SU(2)_2$ to the Heisenberg point at $\theta = 0$, described by a sigma model without topological term \cite{2books}.
Since both a relevant operator and a marginal one appear in (\ref{HdimHal}), then a simple power law behavior for the mass gap and the magnetization is not expected even in the infinite limit, but a logarithmic correction to it. This picture was confirmed in \cite{Marston}. 
Closely related, for a finite chain a marginal operator has a dramatic effect on the scaling dimensions of the operators of the infrared continuos theory, as argued in \cite{Affshort}. In particular, one can define effective scaling dimensions which flow logarithmically slowly in $L$ to their asymptotic values. For the dimer order parameter and for a chain having 256 sites a scaling dimension $x  = 0.48$ was measured in \cite{Affshort}, to be compared with the infinite $L$ value $x = 3/8 = 0.375$. The value 0.48 is very close to the value $x = 0.47$ that we obtained for a chain of the same length $L = 256$  using  the relation $x= \frac{\beta}{\nu}$ and the measured values $\beta=0.66$ and $\nu=1.41$.

\subsection{Large D - Haldane phase transition}
We consider now the transition between the large D phase and the Haldane phase. This phase transition line is described by a CFT with central charge $c  = 1$ related to  a free boson compactified on a circle \cite{gaussian}. The boson velocity, the radius of the circle and importantly the operator content of the theory depend on the parameters $\theta$ or $D$.  This fact implies that the critical exponents are continuously varying changing the transition point along the critical line \cite{gaussian}.  We focus on the transition at $\theta=0$ (see arrow  $3$ in Fig.~\ref{fig:phase}). Our results for the Schmidt gap provide
the critical point at $D^{\Delta\lambda}_c = 0.96$ and $\nu^{\Delta\lambda}=1.56$. These results are again compatible from those obtained with Quantum MonteCarlo~\cite{Albuquerque09} simulations: $D^{QM}_c=0.971$ and $\nu^{QM}=1.4$.  Also in this case  the scaling law ansatz (\ref{powerfit}) looks to be pretty accurate and provides a critical exponent $\beta_{\Delta\lambda}=0.3$. Again the Schmidt gap turns out to be a reliable order parameter.

\subsection{N\'eel - dimer phase transition}
For completeness we investigate now the QPT between the
dimer and the N\'eel phases (arrow $4$ in Fig.~\ref{fig:phase}) and analyze the scaling
of the Schmidt gap along this transition. Our DMRG calculations of the entanglement entropy $S$ show that, decreasing the parameter $D$ from $D = 0$ to $D = - \infty$, the central charge along the transition line varies continuously from $c = \frac{3}{2}$ to $c =1$. In order to extract $c$ we used the formula $S=\frac{c}{6} \,  \mathrm{Log} \, L$ describing the scaling of the half chain entanglement entropy at criticality \cite{Holzhey94,Calabrese04}. The observed peculiar behavior
it is possible because the (1+1)D conformal field theories have a continuos set of central charges and Verma modules for $c \geq 1$ \cite{Cardy,Dif}.\\
Also in this case we expect the Schmidt gap to close at the critical point. Indeed, for $D=-0.9$, $\Delta\lambda$ exhibits critical scaling around $\theta=-0.45\pi$ with critical exponents $\beta^{\Delta\lambda}=0.19$ and $\nu=1.2$ while the FSS analysis of the dimer order parameter leads to the same critical point $\theta=-0.45\pi$ and to the exponents  $\nu = 1.0$ and $\beta_{\mathcal{D}} = 0.75$.

\section{Discussion and conclusions}
\label{sec:conclusion}
Here we have analyzed different QPTs, either integrable or not, by the scaling properties of the Schmidt gap,
 i.e. the difference between the two largest non degenerate Schmidt eigenvalues \cite{DeChiara_Lepori}.
We have derived analytically the scaling of the Schmidt gap for the spin-$\frac{1}{2}$ Ising chain in a transverse or a longitudinal magnetic field and we have checked the effectiveness of these formulae on the numerical data obtained by direct diagonalization or DMRG respectively. This succesful comparison, especially in the case of the transverse Ising model, supports the old conjecture by Baxter \cite{Baxter}  and Cardy \cite{Cardy} relating the Schmidt eigenvalues close to a phase transition to the characters of the conformal theory at the critical point.\\
By means of DMRG simulations, we have studied  the Schmidt gap for various phase transitions of the bilinear-biquadratic spin-1 chain with a quadratic Zeeman term. This study allows us to extend some of our results to a wider set of QPTs than the transitions of the two Ising universality classes, described for instance by non-integrable infrared effective models or characterized by the presence of continuos symmetries. \\
In particular, our findings have shown that the Schmidt gap correctly signals a QPT and, approaching the critical point, it displays a scaling behavior as a conventional local order parameter. This scaling is characterized by critical exponents closely related to the CFT describing the critical point and to its specific perturbations. Moreover, thanks to the power of the FSS technique, the mass scaling exponent $\nu$ turns out to be computable with good precision from the study of the Schmidt gap.  This allows us to identify the Schmidt gap as a non-local order parameter, particularly valuable whenever a ``standard" local order parameter is not known, for instance because it cannot be defined at all, as in the case of topological phases, or because the nature of the phase itself is obscure.\\
At first glance, the effectiveness of the Schmidt gap to locate phase transitions would stem from the behavior of the entanglement spectrum, that at criticality and in the  thermodynamical limit must tend to a continuos distribution: this simply implies the closure of the Schmidt gap. Although finite size effects generally prevent this closure, the scaling behavior survives and it can be recovered by FSS analysis.\\
The analytic derivation of the Schmidt gap scaling for transitions in the Ising universality classes suggests that a scaling behavior should also hold for the difference between higher order Schmidt eigenvalues $|\lambda_i - \lambda_j|$, however in the numerical data  from exact diagonalization (for the transverse Ising model) and form DMRG no scaling and not even closing is observed for them.  
We leave as an open point whether this lack of scaling is only due to finite size effects.\\
Remarkably, one result of the present paper is also a direct check of the results in \cite{Wen11,Cirac11} about the classification of topological order in short-range interacting gapped quantum many body systems: the analysis of the entanglement spectrum along the parameter line $D = 0$, including the vicinity of the dimer-Haldane phase transition, displays a multiplet structure with statistics in agreement with these results.\\
The observed properties of the entanglement spectrum and of the Schmidt gap can be an important boost for future developments concerning the characterization of strongly interacting systems with higher dimensionality/ and or with long range and order, as in
ultracold polar molecules systems \cite{dipolar} or in presence of deconfined quantum critical points \cite{Senth04,Bern04}, where moreover a peculiar entanglement structure has been recently found \cite{SWSE11}.  Indeed,  first results showing scaling properties of the entanglement spectrum near QPT for 2D quantum systems have very recently appeared\cite{Lauchli13,James12}. The analysis we propose is particularly relevant for fermionic systems, when the sign problem forbids trustable numerical results obtained via a MonteCarlo approach. Some preliminary attempts to describe long-range interacting systems in one dimension are already available \cite{Nie12}. A central topic along this direction is the classification of the topological order and phases in such systems, enlarging  the logical scheme adopted in  \cite{Wen11,Cirac11}. The entanglement spectrum has been shown to encode important features of topological order and its analysis,  along the lines of the present paper, is expected to have a primary importance.\\

{\it Acknowledgements:} We thank P. Calabrese, G. Delfino and M. Lewenstein
for enlightening discussions. We acknowledge financial support from the
Spanish MINECO (FIS2008-01236), European Regional development Fund, Generalitat de Catalunya Grant
No. SGR2009-00347, UAB post-doc fellowship awarded by Banco de Santander. 


\begin{thebibliography}{99}
\bibitem{note2}
For an infinite system with translational invariance the location of the partition cut does not affect the results.
However, for finite systems the location of the cut gives corrections to entanglement entropy and entanglement spectrum. 
One has to be careful for instance in presence of a symmetry-breaking for the cut,
as translation invariance in a dimerized system or in the presence of a charge density wave \cite{deng}.
\end{thebibliography}
\end{document}